\documentclass[10pt,conference]{IEEEtran}
 \IEEEoverridecommandlockouts

\usepackage{array}
\usepackage{mdwmath}
\usepackage{mathtools}
\usepackage{subfig}
\usepackage{graphicx}
\usepackage{url}
\usepackage[noend]{algpseudocode}
\usepackage[ruled,vlined]{algorithm2e}
\graphicspath{{../pdf/}{../jpeg/}}
\DeclareGraphicsExtensions{.pdf,.jpeg,.png}

\usepackage{amsfonts}
\usepackage{booktabs}
\usepackage{lipsum}
\usepackage{tikz}
\usetikzlibrary{%
patterns,%
calc,%
fit,%
arrows,%
plotmarks,%
shadows,%
chains,%
shapes%
}
\usepackage{pgfplots}
\usepackage{cite}
\usepackage{amsmath,amssymb,amsfonts}
\newtheorem{lemma}{Lemma}
\newtheorem{theorem}{Theorem}

\usepackage{latexsym}
\usepackage{amsfonts,amssymb,amsmath}

\begin{document}

\title{Enhancing Secret Key Generation in Block Fading Channels using Reconfigurable Intelligent Surfaces}
\author{
\IEEEauthorblockN{Hibatallah Alwazani, \IEEEmembership{Student Member, IEEE}, Anas Chaaban, \IEEEmembership{Senior Member, IEEE}}\\

\thanks{%
The authors are with the School of Engineering, the University of British Columbia, 1137 Alumni Ave., Kelowna, BC V1V1V7, Canada (email: \{hibat97,anas.chaaban\}@ubc.ca)}
}
\maketitle
\begin{abstract}


Physical-layer security (PLS) is superior to classical cryptography techniques due to its notion of perfect secrecy and independence to an eavesdropper's computational power. One form of PLS arises when Alice and Bob (the legitimate users) exchange signals to extract a common key from the random common channels. The drawback of extracting keys from wireless channels is the ample dependence on the dynamicity and fluctuations of the radio channel. However, some radio channels are constant such as line-of-sight (LoS) and can be estimated by Eve  (an illegitimate user), or can be quite static in behavior due to the presence low-mobility users thus restricting the amount of randomness. This in turn lowers the secret key rate (SKR) defined as the number of bits of key generated per channel use. In this work, we aim to address this challenge by using a reconfigurable intelligent surface (RIS) to produce random phases at certain carefully curated intervals such that it disrupts the channel in low-entropy environments. 
We propose an RIS assisted key generation method, study its performance, and compare with benchmarks to observe the benefit of using an RIS while considering various important metrics such as key mismatch rate and average secret key throughput. Simulations are made to validate our theoretical findings showing an improvement in performance when an RIS is deployed.  
\end{abstract}


\section{Introduction}

\label{intro}
Physical layer security (PLS) provides means for information transmission that is provably secure. It is more desirable than classical cryptography \cite{cryptobook} for several reasons \cite{maurer}. First, there is no assumption made on the eavesdropper's computational power, they could have, in principle, unlimited computation power and still not be able to decipher the message in some scenarios. Second, it gives rise to the notion of perfect secrecy, where knowing the ciphertext at an eavesdropper tells it nothing about the message being exchanged. {Finally, it is highly scalable \cite{plsRenzo}, which is necessary for future generation networks as devices connected to the nodes may have varying power and computation capabilities.}

\begin{figure}
    \centering
    \tikzset{every picture/.style={scale=1}, every node/.style={scale=1}}
    \begin{tikzpicture}
    \node (a) at (0,0) {};
    \node at ($(a)+(0,-.1)$) {\footnotesize Alice};
    \draw[line width=1] ($(a)+(-.2,.1666)$) to ($(a)+(0,.75)$) to ($(a)+(.2,.1666)$);
    \draw[line width=1] ($(a)+(.2,.1666)$) to ($(a)+(-.12,.3666)$);
    \draw[line width=1] ($(a)+(-.2,.1666)$) to ($(a)+(.12,.3666)$);

    \draw[line width=1] ($(a)+(.08,.5666)$) to ($(a)+(-.12,.3666)$);
    \draw[line width=1] ($(a)+(-.082,.5666)$) to ($(a)+(.12,.3666)$);
    \draw[line width=1.5] ($(a)+(-.09,.75)$) to  ($(a)+(.09,.75)$);
    \draw[line width=1.5] ($(a)+(0,.75)$) to  ($(a)+(0,.8)$);
    
	\node (b) at (6,-.1) {\ \ \footnotesize Bob};
	\draw[fill=gray!50!white] ($(b)+(0,.3)$) to ($(b)+(.25,.3)$) to ($(b)+(.3,.8)$) to ($(b)+(.05,.8)$) to ($(b)+(0,.3)$); 
	
	\node (e) at (3,-1.6) {\ \ \footnotesize Eve};
	\draw[fill=gray!50!red] ($(e)+(0,.3)$) to ($(e)+(.25,.3)$) to ($(e)+(.3,.8)$) to ($(e)+(.05,.8)$) to ($(e)+(0,.3)$); 
	    
	\node (r) at (3,3) {\footnotesize Rose};
	\node (r_corner) at ($(r)-(.75,.2)$) {};
    \draw[blue!50!white] (r_corner.center) to ($(r_corner.center)+(1.5,0)$) to 
    ($(r_corner.center)+(1.5,-1.5)$) to ($(r_corner.center)-(0,1.5)$) to (r_corner.center);
    \foreach \i in {1,..., 9}
    {
    \draw[blue!50!white] ($(r_corner.center)+(.15*\i,0)$) to ($(r_corner.center)+(.15*\i,-1.5)$);
    \draw[blue!50!white] ($(r_corner.center)+(0,-.15*\i)$) to ($(r_corner.center)+(1.5,-.15*\i)$);
    }
    
    \draw[->] ($(a)+(.2,.7)$) to node [above,sloped] {\footnotesize $\mathbf{h}_{\rm ar}$} ($(r)-(0,.95)$);
    \draw[->,dashed] ($(r)-(0,1.1)$) to node [below,sloped] {\footnotesize $\mathbf{h}_{\rm ra}$} ($(a)+(.2,.55)$);
    
    \draw[->,dashed] ($(b)+(-.2,.55)$) to node [below,sloped] {\footnotesize $\mathbf{h}_{\rm br}$} ($(r)-(-0.1,1.1)$);
    \draw[->] ($(r)-(-0.1,.95)$) to node [above,sloped] {\footnotesize $\mathbf{h}_{\rm rb}$} ($(b)+(-.2,.7)$);

    \draw[<-] ($(b)+(-.3,.45)$) to node [above,sloped] {\footnotesize $h_{\rm ab} \qquad \qquad $} ($(a)+(0.3,.45)$);
    \draw[<-,dashed] ($(a)+(0.3,.3)$) to node [below,sloped] {\footnotesize $h_{\rm ba} \qquad \qquad $} ($(b)+(-.3,.3)$);

    \draw[->] ($(a)+(0.3,.1)$) to node [below,sloped] {\footnotesize $h_{\rm ae}$} ($(e)+(-.2,.5)$);
    \draw[->,dashed] ($(b)+(-0.3,.1)$) to node [below,sloped] {\footnotesize $h_{\rm be}$} ($(e)+(.4,.5)$);
    \draw[->] ($(r)-(-0.05,1.2)$) to node [above,sloped] {\footnotesize $h_{\rm re}\qquad$} ($(e)+(.1,.9)$);

    \end{tikzpicture}
    \caption{System model with an access point Alice, receiver Bob, RIS Rose, and eavesdropper Eve.}
    \label{fig:sysmodel}
\end{figure}
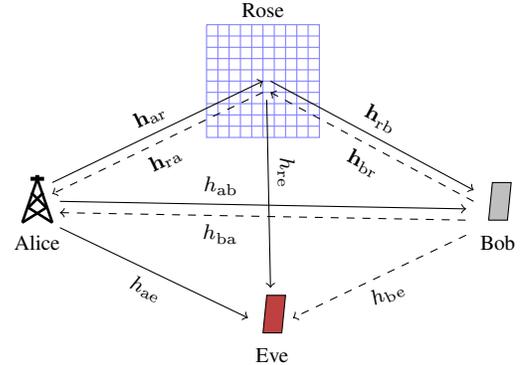
Generally, for key-based secrecy, we face a fundamental question in cryptography: how do two parties share a secret key without compromising the key? In PLS, a key can be extracted from the wireless channel which acts as a common and unique source of randomness for Alice (the transmitter) and Bob (the receiver). This idea of generating random, common, and secret keys at two legitimate points from wireless channels is not new \cite{keyreview,networkinfo}. It has been explored before but largely abandoned  because of the fundamental limitation of low-entropy environments where channels vary slowly thereby producing low secret key rates (SKR) \cite{unsuitableBits}. However, with the emergence of a smart radio environment enabled by the practical implementation of the reconfigurable intelligent surfaces (RIS)s, this limitation may be overcome \cite{smartRadio}. Here, the RIS is abstracted as a large number of passive, scattering elements, where each element can be reconfigured to change the amplitude and/or phase of the impinging electromagnetic (EM) waves to achieve a desired objective, such as inducing channel variations in our case. Essentially, the RIS becomes a building block for a programmable and software-defined wireless environment \cite{softwarecontrolIRS}. Thus, this aspect in physical layer security (PLS) is undergoing a resurgence, with a focus on smart radio environment enabled secret key generation.

The authors in \cite{keygeneration} present a novel wireless key generation architecture based on randomized channel responses from an RIS which act as the shared random source to Alice and Bob. They present their results using two metrics which are SKR and key mismatch rate (KMR). The authors in \cite{multiuser} propose a joint user allocation scheme and an RIS reflection parameter adjustment scheme to enhance key generation efficiency in a multi-user communication scenario. They compare the result against a scheme without an RIS and find that the RIS indeed boosts performance by reducing channel similarities between adjacent users and thus the enhancing efficiency of key generation.

The authors in \cite{chen2021intelligent} derive an upper bound on the SKR using \cite{maurer} and compare with another SKR upper bound without the presence of an RIS in the system. 
The work in \cite{lowerboundSKR} studies the minimum achievable SKR in the presence of an RIS and multiple passive eavesdroppers, where the authors optimize the minimum SKR by choosing appropriate RIS phase shifts. Here, the RIS works in a capacity to combat deleterious wireless channel conditions such as co-channel interference and dead zones. Moreover, \cite{lowerboundSKR} consider an RIS as a new degree of freedom in the channel, where the aim of the RIS is increasing correlation between legitimate nodes' channels and decreasing correlation with eavesdropping channels. However, \cite{lowerboundSKR} assumes that channel state information (CSI) is known at Alice and the RIS which is not practical. 

In this paper, we study a practical secret key generation protocol, where CSI is initially unknown at all nodes. The RIS perturbs the block fading channel to induce more randomness in the channel. Then, channels are estimated and keys are generated over multiple blocks using the perturbed channel which changes faster than the original block fading channel, thanks to the RIS. The contributions of the paper can be summarized as follows:
\begin{itemize}
    \item We formulate a theoretical achievable SKR lower bound for the proposed protocol.
    \item For a practical implementation of the protocol, we study the key mismatch rate (KMR) and the key throughput defined as the average number of key bits generated per transmission.
    \item We study the effect of the RIS in terms of several parameters such as the number of elements and the RIS switching rate.
\end{itemize}
In general, we notice a significant improvement in performance compared to the scenario without an RIS, where the key rate is limited due to the static nature of the channels within a fading block. 
To delve deeper into the results, we start by formulating the system model studied in this paper.

 



\section{System Model}
Consider the setup depicted in Fig. \ref{fig:sysmodel}, where a single antenna access point (Alice) serves a single-antenna user (Bob), in the presence of a passive eavesdropper (Eve) and an RIS (Rose) equipped with $N$ elements controlled by Alice through a control link. To secure the communication, Alice and Bob generate a key through exchanging signals over the wireless channel, and use the key to encrypt the message to be transmitted. Eve knows the cryptosystem and the key generation protocol, and aims to discover the information exchanged between Alice and Bob.


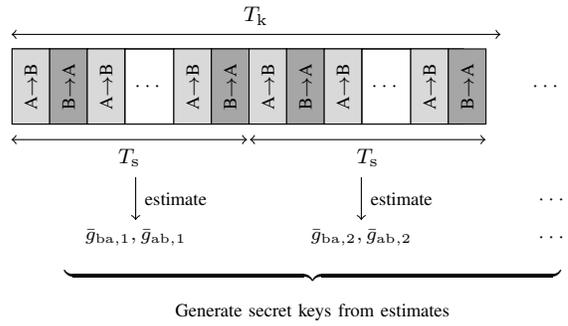
\begin{figure}
    \centering
\begin{tikzpicture}
\draw[<->] (0,1.2) to node[above]{\footnotesize $T_{\rm k}$} (6.5,1.2);
\node[anchor=south west,draw,minimum width = 6cm,minimum height = 1cm] at (0,0) {};

\node[anchor=south west,draw,minimum width = .5cm,minimum height = 1cm,fill=gray!30!white] at (0,0) {\rotatebox{90}{\scriptsize A$\to$B}};
\node[anchor=south west,draw,minimum width = .5cm,minimum height = 1cm,fill=gray!70!white] at (.5,0) {\rotatebox{90}{\scriptsize B$\to$A}};
\node[anchor=south west,draw,minimum width = .5cm,minimum height = 1cm,fill=gray!30!white] at (1,0) {\rotatebox{90}{\scriptsize A$\to$B}};
\node[anchor=south west,draw,minimum width = .65cm,minimum height = 1cm] at (1.5,0) {\scriptsize $\cdots$};
\node[anchor=south west,draw,minimum width = .5cm,minimum height = 1cm,fill=gray!30!white] at (2.15,0) {\rotatebox{90}{\scriptsize A$\to$B}};
\node[anchor=south west,draw,minimum width = .5cm,minimum height = 1cm,fill=gray!70!white] at (2.65,0) {\rotatebox{90}{\scriptsize B$\to$A}};
\draw[<->] (0,-.2) to node[below]{\footnotesize $T_{\rm s}$} (3.13,-.2);

\node[anchor=south west,draw,minimum width = .5cm,minimum height = 1cm,fill=gray!30!white] at (3.15,0) {\rotatebox{90}{\scriptsize A$\to$B}};
\node[anchor=south west,draw,minimum width = .5cm,minimum height = 1cm,fill=gray!70!white] at (3.65,0) {\rotatebox{90}{\scriptsize B$\to$A}};
\node[anchor=south west,draw,minimum width = .5cm,minimum height = 1cm,fill=gray!30!white] at (4.15,0) {\rotatebox{90}{\scriptsize A$\to$B}};
\node[anchor=south west,draw,minimum width = .65cm,minimum height = 1cm] at (4.65,0) {\scriptsize $\cdots$};
\node[anchor=south west,draw,minimum width = .5cm,minimum height = 1cm,fill=gray!30!white] at (5.3,0) {\rotatebox{90}{\scriptsize A$\to$B}};
\node[anchor=south west,draw,minimum width = .5cm,minimum height = 1cm,fill=gray!70!white] at (5.8,0) {\rotatebox{90}{\scriptsize B$\to$A}};
\draw[<->] (3.17,-.2) to node[below]{\footnotesize $T_{\rm s}$} (6.3,-.2);

\node[anchor=south west,minimum width = .5cm,minimum height = 1cm] at (6.8,0) {\scriptsize $\cdots$};

\node (gab1) at (1.65,-1.5) {\scriptsize $\bar{g}_{{\rm ba},1},\bar{g}_{{\rm ab},1}$};
\draw[->] (1.65,-.7) to node[right]{\scriptsize estimate} (gab1);

\node (gab2) at (4.65,-1.5) {\scriptsize $\bar{g}_{{\rm ba},2},\bar{g}_{{\rm ab},2}$};
\draw[->] (4.65,-.7) to node[right]{\scriptsize estimate} (gab2);
\node at (7.2,-1) {\scriptsize $\cdots$};
\node at (7.2,-1.5) {\scriptsize $\cdots$};

\node at (4,-2) {\scriptsize $\underbrace{\hspace{6.6cm}}$};
\node at (4,-2.5) {\scriptsize Generate secret keys from estimates};

\end{tikzpicture}
\caption{Key generation time $T_{\rm k}$ split into multiple RIS switching periods with duration $T_{\rm s}$ from each of which a channel estimate is obtained at Alice (A) and Bob (B). The estimates are then put through a process to generate a common key.}
    \label{fig:coherence}
\end{figure}

As can be seen in Fig. \ref{fig:sysmodel}, we denote by ${h}_{\rm ab},h_{\rm ba}\in\mathbb{C}$ the channels from Alice to Bob and from Bob to Alice, respectively, by $h_{\rm ae},h_{\rm be}\in\mathbb{C}$ the channels from Alice and Bob to Eve, respectively, by $\mathbf{h}_{{\rm ar}},\mathbf{h}_{{\rm br}}  \in \mathbb{C}^{N}$ the channels from Alice and Bob to Rose, respectively, and by $\mathbf{h}_{\rm ra},\mathbf{h}_{\rm rb},\mathbf{h}_{\rm re} \in \mathbb{C}^{N}$ the channels from Rose to Alice, Bob, and Eve, respectively.

We assume block-fading channels in which all channels maintain constant values for $T$ symbols and vary independently between blocks based on their respective distributions. We also assume a time-division duplexing (TDD) scheme, which implies that $h_{\rm ba}=h_{\rm ab}$, $\mathbf{h}_{\rm ra}=(\mathbf{h}_{\rm ar}^H)^T$, and $\mathbf{h}_{\rm rb}=(\mathbf{h}_{\rm br}^H)^T$ (reciprocal channels). Moreover, we assume independent Rayleigh fading so that 
\begin{align}
\label{model_1}
&{h}_{ij}\sim \mathcal{CN}(0, \beta_{ij})\\
&\mathbf{h}_{{\rm r}i}\sim \mathcal{CN}(0, \beta_{{\rm r}i}\mathbf{I}_{N})
\end{align}
for $i,j \in \{{\rm a,b,e}\}$, where $\mathcal{CN}(0, \mathbf{Q})$ denotes a circularly symmetric complex Gaussian distribution with mean $0$ and covariance matrix $\mathbf{Q}$. 

The transmission time is divided between Alice and Bob for the sake of key generation and information transmission as shown in Fig. \ref{fig:coherence}. To realize secure transmission, a transmission block of length $T$ symbols representing a coherence interval is divided into $T_{\rm k}$ symbols used for key generation, and $T_{\rm d}$ symbols used for data transmission. We focus on the key generation phase in this work. During the key generation phase, Alice transmits during odd time slots, while Bob transmits during even time slots. Denoting the transmitted symbols by Alice and Bob by $x_{{\rm a},t}\in\mathbb{C}$ and $x_{{\rm b},t}\in\mathbb{C}$, respectively, which satisfy the power constraints $\sum_{t\text{ odd}} |x_{{\rm a},t}|^2\leq \frac{T_{\rm k}}{2}P$ and $\sum_{t\text{ even}} |x_{{\rm b},t}|^2\leq \frac{T_{\rm k}}{2}P$, the received signals can be written as
\begin{align}
{y}_{i,t} &=  (h_{{\rm a}i} + \mathbf{h}_{\rm ar}^H \boldsymbol{\Phi}_t \mathbf{h}_{{\rm r}i} ) x_{{\rm a},t} + n_{i,t}, \ \  i \in \{{\rm b, e}\},\ t \text{ odd,}\label{RxSignalBobandEve}\\
{y}_{i,t} &=  (h_{{\rm b}i} + \mathbf{h}_{\rm br}^H \boldsymbol{\Phi}_t \mathbf{h}_{{\rm r}i} ) x_{{\rm b},t} + n_{i,t}, \ \ i \in \{{\rm a, e}\},\ t \text{ even,}
\end{align}
where  
\begin{align}
\label{Phi_def}
\boldsymbol{\Phi}_t=\text{diag}([e^{j \theta_{t,1}}, \dots,e^{j\theta_{t,N}}])\in \mathbb{C}^{N\times N},
\end{align}
is the reflection matrix for Rose in time slot $t$, $\theta_{t,n}\in [0,2\pi]$ is the random phase-shift applied by element $n$, and $n_{{\rm a},t},n_{{\rm b},t},n_{{\rm e},t}\in\mathbb{C}$ are noise samples at Alice, Bob, and Eve, respectively, which are independent of each other, and are independent and identically distributed over time with distribution $\mathcal{CN}(0,\sigma^2)$.
Using this transmission, Alice and Bob can generate a shared key $\mathbf{k}=(k_1,\ldots,k_r)\in\{0,1\}^r$ where $r$ is the total number of key bits. {Since $T_k$ is defined as the total number of symbols reserved for key generation at the two nodes,  Alice and Bob split this portion in half  for their respective key generation as $\frac{T_k}{2}$}. Thus,  the secret key rate (SKR) in bits per symbol is defined as $R_{\rm k}=\frac{r}{T_{\rm k}/2}$, which is desired to be large.
The end goal of Alice and Bob is to perform this key generation and extract $\mathbf{k}$ while preventing Eve from being able to discover the key.

\section{Key Generation and Secret Key Rate}
Alice and Bob use the random channel between them as a source of shared randomness to generate a key. Since the channel remains constant during a coherence interval of length $T$ symbols, the RIS can help disrupt the channel by embedding additional randomness during a coherence interval. \cite{RRS} introduces the concept of random reconfigurable surfaces (RRS) accounting for RISs that whose elements induce a time-variant phase shift on the reflected signals and present it as the diffusion function of an RIS. This is the same functionality we use for the RIS implementation detailed next.

\subsection{Channel Estimation}
Alice and Bob generate keys by estimating their channels (Fig. \ref{fig:coherence}) and using the channel estimates as common randomness. To randomize the channel during a coherence interval, we consider an RIS which switches its phase-shift matrix $\boldsymbol{\Phi}_t$ randomly every $T_{\rm s}$ symbols, with $T_{\rm s}\leq T_{\rm k}$ and $\theta_{t,n}\sim\text{Uniform}[0,2\pi]$. Let the RIS phase shift matrix during switching period $\ell$ be represented by $\boldsymbol{\Phi}_{\ell}$. Alice and Bob estimate the channel for each switching interval $\ell\in\{1,\ldots,\frac{T_{\rm k}}{T_{\rm s}}\}$. Alice and Bob send pilot signals $\mathbf{x}_{{\rm a},\ell},\mathbf{x}_{{\rm b},\ell}\in\mathbb{C}^{T_{\rm s}/2}$ in switching period $\ell$ during odd-indexed and even-indexed symbols, respectively, such that $\|\mathbf{x}_{{\rm a},\ell}\|^2=\|\mathbf{x}_{{\rm a},\ell}\|^2=\frac{T_{\rm s}}{2}P$. Alice and Bob receive
\begin{align}
\mathbf{y}_{{\rm a},\ell}&=g_{\rm ba}\mathbf{x}_{{\rm b},\ell}+\mathbf{n}_{{\rm a},\ell},\\
\mathbf{y}_{{\rm b},\ell}&=g_{\rm ab}\mathbf{x}_{{\rm a},\ell}+\mathbf{n}_{{\rm b},\ell},
\end{align}
where $g_{{\rm ba},\ell}=h_{\rm ba} + \mathbf{h}_{\rm br}^H \boldsymbol{\Phi}_{\ell} \mathbf{h}_{\rm ra}$ and $g_{{\rm ab},\ell}=h_{\rm ab} + \mathbf{h}_{\rm ar}^H \boldsymbol{\Phi}_{\ell} \mathbf{h}_{\rm rb}=g_{{\rm ba},\ell}$, and $\mathbf{n}_{{\rm a},\ell}$ and $\mathbf{n}_{{\rm b},\ell}$ collect the noise instances during switching period $\ell$ during odd-indexed and even -indexed symbols, respectively. Alice and Bob then estimate the channels $g_{{\rm ba},\ell}$ and $g_{{\rm ab},\ell}$ to be used as shared randomness as follows (using least-squares estimation)
\begin{align}
\bar{g}_{{\rm ba},\ell}&=\mathbf{y}_{{\rm a},\ell}^H\frac{\mathbf{x}_{{\rm b},\ell}}{\|\mathbf{x}_{{\rm b},\ell}\|^2} = g_{{\rm ba},\ell} + \bar{n}_{{\rm ba},\ell},\label{Estimate_ba}\\
\bar{g}_{{\rm ab},\ell}&=\mathbf{y}_{{\rm b},\ell}^H\frac{\mathbf{x}_{{\rm a},\ell}}{\|\mathbf{x}_{{\rm a},\ell}\|^2} = g_{{\rm ab},\ell} + \bar{n}_{{\rm ab},\ell},
\end{align}
where $\bar{n}_{{\rm ba},\ell}=\frac{\mathbf{n}_{{\rm a},\ell}^H\mathbf{x}_{{\rm b},\ell}}{\|\mathbf{x}_{{\rm b},\ell}\|^2}$ and $\bar{n}_{{\rm ab},\ell}=\frac{\mathbf{n}_{{\rm b},\ell}^H\mathbf{x}_{{\rm a},\ell}}{\|\mathbf{x}_{{\rm a},\ell}\|^2}$ are independent $\mathcal{CN}(0,\bar{\sigma}^2)$ noises with $\bar{\sigma}^2=\frac{2\sigma^2}{T_{\rm s}P}$. During the same time, Eve obtains the following estimates similarly
\begin{align}
\bar{g}_{{\rm be},\ell}&= g_{{\rm be},\ell} + \bar{n}_{{\rm be},\ell},\label{Estimate_be}\\
\bar{g}_{{\rm ae},\ell}&= g_{{\rm ae},\ell} + \bar{n}_{{\rm ae},\ell},
\end{align}
where $g_{{\rm be},\ell}=h_{\rm be} + \mathbf{h}_{\rm br}^H \boldsymbol{\Phi}_{\ell} \mathbf{h}_{\rm re}$, $g_{{\rm ae},\ell}=h_{\rm ae} + \mathbf{h}_{\rm ar}^H \boldsymbol{\Phi}_{\ell} \mathbf{h}_{\rm re}$, and $\bar{n}_{{\rm be},\ell}$ and $\bar{n}_{{\rm ae},\ell}$ are independent $\mathcal{CN}(0,\bar{\sigma}^2)$ noises.

Next, Alice and Bob use the estimates $\bar{g}_{{\rm ba},\ell}$ and $\bar{g}_{{\rm ab},\ell}$ (which are dependent) to generate their shared key (as in \cite[Ch. 22]{elgamal_kim}). Note that since the channels are assumed to not vary during a coherence block and to vary between blocks, the channel estimates will not be independently and identically distributed (i.i.d.) across multiple blocks. However, Alice and Bob can use the $\ell^{\rm th}$ switching period in multiple coherence blocks to generate one key, making the estimates used to generate a given key i.i.d. as desired. This means that $\frac{T_{\rm k}}{T_{\rm s}}$ keys can be generated simultaneously. Alternatively, we can use interleaving to obtain a pseudo-i.i.d. estimates by taking the estimates from the first switching periods from a set of blocks, followed by the estimates from the second switching period, and so on. Next, we characterize the SKR.

\subsection{Secret Key Rate}
Let the estimates $\bar{g}_{{\rm ba},\ell}$, $\bar{g}_{{\rm ab},\ell}$, $\bar{g}_{{\rm be},\ell}$ and $\bar{g}_{{\rm ae},\ell}$ be represented by random variables $\bar{G}_{{\rm ba}}$, $\bar{G}_{{\rm ab}}$, $\bar{G}_{{\rm be}}$ and $\bar{G}_{{\rm ae}}$, respectively. The SKR can be lower bounded by 
$$R_{\rm k}\geq  \frac{1}{T_{\rm s}/2} R_{\rm k}^{\rm lb},$$ where \cite{maurer} 
\begin{align}\label{SKR_lb}
R_{\rm k}^{\rm lb}&\triangleq I(\bar{G}_{\rm ab}; \bar{G}_{\rm ba})\\
&\qquad-\min\{I(\bar{G}_{\rm ab};\bar{G}_{\rm ae},\bar{G}_{\rm be}),I(\bar{G}_{\rm ba};\bar{G}_{\rm ae},\bar{G}_{\rm be})\}\nonumber,\\
&=-{\sf h}(\bar{G}_{\rm ab}| \bar{G}_{\rm ba})\nonumber\\
&\qquad+\max\{{\sf h}(\bar{G}_{\rm ab}|\bar{G}_{\rm ae},\bar{G}_{\rm be}),{\sf h}(\bar{G}_{\rm ba}|\bar{G}_{\rm ae},\bar{G}_{\rm be})\},\nonumber
\end{align}
$I(X;Y)$ is the mutual information, ${\sf h}(X|Y)$ is the conditional entropy, and the factor $\frac{1}{T_{\rm s}/2}$ follows because there are $\frac{T_{\rm k}}{T_{\rm s}}$ channel estimates in key generation phase of duration $T_{\rm k}$. To simplify this lower bound, we need to study the distributions of the channel estimates. 
We first note that $\bar{G}_{{\rm ba}}$, $\bar{G}_{{\rm ab}}$, $\bar{G}_{{\rm be}}$ and $\bar{G}_{{\rm ae}}$ have zero mean. Moreover, the covariances of the channels are given by
\begin{align}
\rho_{\rm ab}&=\mathbb{E}[G_{\rm ab}G_{\rm ab}^*]=\mathbb{E}[G_{\rm ab}G_{\rm ba}^*]=\mathbb{E}[G_{\rm ba}G_{\rm ba}^*]\nonumber\\
&=\beta_{\rm ab}+N\beta_{\rm ar}\beta_{\rm rb},\\
\rho_{\rm ae}&=\mathbb{E}[G_{\rm ae}G_{\rm ae}^*]=\beta_{\rm ae}+N\beta_{\rm ar}\beta_{\rm re},\\
\rho_{\rm be}&=\mathbb{E}[G_{\rm be}G_{\rm be}^*]=\beta_{\rm be}^2+N\beta_{\rm br}\beta_{\rm re},
\end{align}
Then, 
\begin{align}
{\rm Cov}\left([\bar{G}_{\rm ab}\ \bar{G}_{\rm ba}]^T\right)
&=\begin{bmatrix} \rho_{\rm ab}+\bar{\sigma}^2 & \rho_{\rm ab}\\
\rho_{\rm ab} & \rho_{\rm ab}+\bar{\sigma}^2\end{bmatrix},\label{Covab}\\
{\rm Cov}\left([\bar{G}_{\rm ab}\ \bar{G}_{\rm ae} \ \bar{G}_{\rm be}]^T\right)&={\rm Cov}\left([\bar{G}_{\rm ba}\ \bar{G}_{\rm ae} \ \bar{G}_{\rm be}]^T\right)\label{Covabe}\\
&\hspace{-1cm}=\begin{bmatrix}
\rho_{\rm ab}+\bar{\sigma}^2 & 0 & 0\\
0 & \rho_{\rm ae}+\bar{\sigma}^2 & 0\\
0 & 0 & \rho_{\rm be}+\bar{\sigma}^2
\end{bmatrix}.\nonumber
\end{align}
Finally, we need the following statement to characterize the distribution of the estimates. 

\begin{lemma}\label{LemmaGaussian}
Given circularly symmetric complex Gaussian channels, the aggregate channel $g_{{ij},\ell}=h_{ij}+\mathbf{h}_{i{\rm r}}^H\boldsymbol{\Phi}_{\ell}\mathbf{h}_{{\rm r}j}$, $i\in\{{\rm a,b}\}$, $j\in\{{\rm a,b,e}\}$, $j\neq i$, can be modeled as a circularly symmetric complex Gaussian when $N$ is large.
\end{lemma}
\begin{IEEEproof}
The statement is a consequence of the central limit theorem, and can be proved similar to \cite[Lemma 2]{impactofPhaseRandomIRS}.
\end{IEEEproof}

Based on this, the lower bound $R_{\rm k}^{\rm lb}$ can be simplified as follows.

\begin{theorem} 
The SKR lower bound in \eqref{SKR_lb} simplifies to
\begin{align}
\label{rkmlb}
R_{\rm k}^{\rm lb}=\frac{1}{T_{\rm s}/2}\log_2\left(1+\frac{\rho^2_{\rm ab}}{\bar{\sigma}^2(2\rho_{\rm ab}+\bar{\sigma}^2)}\right).
\end{align}
\end{theorem}
\begin{IEEEproof}
The statement is obtained by evaluating \eqref{SKR_lb} with circularly symmetric complex Gaussian channel estimates (using Lemma \ref{LemmaGaussian}) and the covariance matrices in \eqref{Covab} and \eqref{Covabe}.
\end{IEEEproof}

Next, we present a protocol for secret key generation using this system.

\section{Protocol for Secret Key Generation}

In this section, we introduce a novel method for creating keys using several coherence blocks.
Over several blocks $f=1,\ldots, F$, we create $\frac{T_{\rm k}}{T_{\rm s}}$ keys. The number of estimates to be used for each key is $F$. In the $\ell^{th}$ switching period in multiple $F$ blocks, Alice and Bob generate keys $\mathbf{k}_{\rm a, \ell}$ and $\mathbf{k}_{\rm b, \ell}$, respectively.  If $\mathbf{k}_{\rm a, \ell}=\mathbf{k}_{\rm b,\ell}=\mathbf{k}_{\ell}$, i.e., the keys match, they are used for encryption. Otherwise, key $\ell$ is discarded. In general, some of the $\frac{T_{\rm k}}{T_{\rm s}}$ keys generated during $F$ blocks will match and will be used for encryption. The process is repeated every $F$ blocks. 
\subsection{Quantization of Channel Estimates} 
Following \cite{secureKeysMultipath}, Alice generates key bits from each block $f$  by quantizing the phase of its channel estimate $\bar{g}_{\rm ba,\ell}$ given by
\begin{align}
    \theta_{\rm ba,\ell}&=\tan^{-1}\bigg(\frac{\text{imag}( \bar{g}_{\rm ba,\ell})}{\text{real}( \bar{g}_{\rm ba,\ell})}\bigg), \ell=1,\dots ,\frac{T_{\rm k}}{T_{\rm s}}.
\end{align}
Bob also does the same using the estimate $\bar{g}_{\rm ab,\ell}$. This happens over the several blocks $F$ until we obtain all the phases of all channel estimates.
We define the quantization of the phase using a function $f_Q:\mathbb{R} \rightarrow\{1,\dots, Q\}$, where $Q$ is the number of quantization levels, such that
\begin{align}
\label{quantizedphase}
 \theta_{ij,\ell}^Q=f_Q( \theta_{ij,\ell})=q, \text{ if } \theta_{ij,\ell}\in \left(\frac{2\pi(q-1)}{Q},\frac{2\pi(q)}{Q}\right),
 \end{align}
for $q=1,\ldots, Q$, $i\in\{\rm a,b\}$ with $i\neq j$. Thus, one channel estimate generates a random phase value that yields $\log_2(Q)$ key bits. The total number of key bits in key $\ell$ denoted by $L$ is thus
\begin{align}
    L={F}\log_2(Q)\text{ bits}.
\end{align} 
Note that a larger $Q$ increases the number of key bits at the expense of higher mismatch probability as discussed next.

 \subsection{Key Mismatch Rate and Throughput Analysis}
For each of the $\frac{T_{\rm k}}{T_{\rm s}}$ keys, we can define key mismatch rate (KMR) as
\begin{align}
    P(\mathbf{k}_{\rm a,\ell}\neq \mathbf{k}_{\rm b,\ell})=1-p,\ \ell=1,\dots ,\frac{T_{\rm k}}{T_{\rm s}},
\end{align} 
where $p$ is the probability that two keys match and amounts to
\begin{align}
\label{matchprob}
    p= [P(\theta_{\rm ba,\ell}^Q=\theta_{\rm ab,\ell}^Q)]^F
\end{align} 
in which $\theta_{\rm ba,\ell}^Q$ and $\theta_{\rm ab,\ell}^Q$ are as defined in \eqref{quantizedphase}.\footnote{After generating a key (quantization), information reconciliation takes place \cite{keyreview} in order to detect and resolve key mismatch scenarios.}
After every key extraction round which extends over $F$ blocks, Alice and Bob check whether they have matching keys. If not, the key is discarded, so that only matching keys are used for encryption. Thus, the number of trials $n$ until a key match occurs for key $\ell$ can be modelled as geometrically distributed random variable $X$ such that
\begin{align}
    P(X=n)=(1-p)^{n-1}p,\ n=1,2,3, \dots
\end{align}
The average number of handshakes $\bar{n}$ until success is given as 
\begin{align}
   \bar{n}= \mathbb{E}[X]=\sum_{n=1}^\infty (1-p)^{n-1}n=\frac{1}{p}.
\end{align}
Then, the average key throughput $\bar{R}_{\rm k}$ can be found as the length of key $\ell$ which is $F\log_2(Q)$ multiplied by the number of keys in $F$ frames which is $\frac{T_{\rm k}}{T_{\rm s}}$, divided by the total symbols allocated per node (Alice and Bob) $F\frac{T_{\rm k}}{2}$ divided by the average number of handshakes $\frac{1}{p}$, yielding
\begin{align}
\label{avgthrough}
    \bar{R}_{\rm k}=\frac{p\log_2(Q)}{T_{\rm s}/2} \text{ bits per symbol.} 
\end{align}

Next, we show simulations that show the effect of RIS on the key generation in terms of number of elements $N$ and switching period $T_s$.


\begin{table}[t]
\vspace{.3cm}
 \begin{center}
\begin{tabular}{c || c}  \hline
Parameter & Value  \\ 
 \hline\hline
 $T_k$ & $40$ symbols   \\ 
 \hline
 $F$ & $100$ blocks \\
 \hline
Noise level & $30$ \rm{dBm} \\
 \hline
$P$ & $1 W$\\
 \hline
$\beta_{ab}=\beta_{ae}=\beta_{be}$& 1\\
 \hline
$\beta_{ar}=\beta_{rb}=\beta_{re}$& 0.7\\
 \hline
\end{tabular}
\end{center}
  \caption{Simulation Parameters.}
    \label{tabsims}
\end{table}

\section{Simulations and Results}
To evaluate the performance of the proposed protocol, we simulate it for a system with parameters provided in Table \ref{tabsims}.    

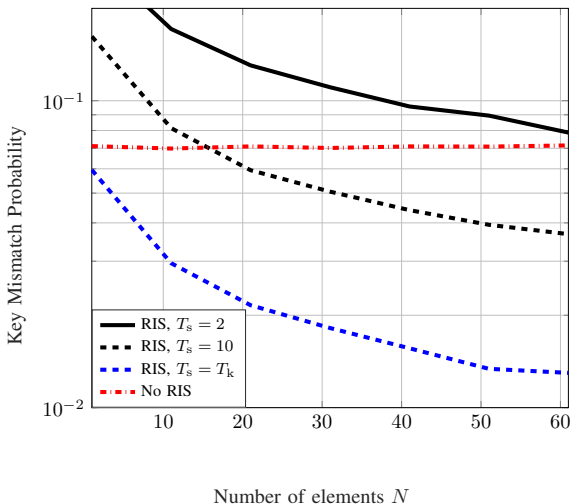
\begin{figure}[t]
\tikzset{every picture/.style={scale=0.77}, every node/.style={scale=1}}
%
%
\begin{tikzpicture}

\begin{axis}[%
width=4.83in,
height=4.15in,
xmin=1,
xmax=61,
xlabel style={at={(0.6,-0.6)}, font=\color{white!15!black}},
xlabel={Number of elements $N$},
ymode=log,
ymin=0.01,
ymax=0.2,
yminorticks=true,
ylabel style={at={(-0.01,0.1)}, font=\color{white!15!black}},
ylabel={Key Mismatch Probability},
xmajorgrids,
ymajorgrids,
yminorgrids,
legend style={at={(axis cs: 1,0.01)}, anchor=south west, legend cell align=left, align=left, draw=white!15!black}
]
\addplot [color=black, line width=2.0pt]
  table[row sep=crcr]{%
1	0.298700000000001\\
11	0.17127\\
21	0.13038\\
31	0.11066\\
41	0.0957899999999998\\
51	0.0894399999999997\\
61	0.0787699999999998\\
};
\addlegendentry{\footnotesize RIS, $T_{\rm s}=2$}

\addplot [color=black, dashed, line width=2.0pt]
  table[row sep=crcr]{%
1	0.16204\\
11	0.0814099999999998\\
21	0.0593700000000001\\
31	0.05056\\
41	0.04401\\
51	0.03943\\
61	0.03676\\
};
\addlegendentry{\footnotesize RIS, $T_{\rm s}=10$}

\addplot [color=blue, dashed, line width=2.0pt]
  table[row sep=crcr]{%
1	0.05953\\
11	0.02955\\
21	0.02156\\
31	0.0181699999999999\\
41	0.0156099999999998\\
51	0.0133799999999998\\
61	0.0129799999999998\\
};
\addlegendentry{\footnotesize RIS, $T_{\rm s}=T_{\rm k}$}

\addplot [color=red, dashdotted, line width=2.0pt]
  table[row sep=crcr]{%
1	0.07115\\
11	0.0698700000000001\\
21	0.071\\
31	0.0701599999999999\\
41	0.071\\
51	0.0709\\
61	0.0715199999999999\\
};
\addlegendentry{\footnotesize No RIS}

\end{axis}
\end{tikzpicture}%
    \caption{Key mismatch rate versus number of RIS elements $N$ for different switching rates of the RIS as well as a no RIS scenario.}
    \label{fig:kmr}
\end{figure}

We start by evaluating the KMR of the protocol. Fig. \ref{fig:kmr} shows the KMR of the protocol versus $N$ for different scenarios: a system without an RIS in which the channels only consist of direct channels, i.e. $g_{{\rm ba},\ell}=h_{\rm ba}$ and $g_{{\rm ab},\ell}=h_{\rm ab} $, and a system with an RIS without switching where $T_s=T_k$, RIS with switching where $T_s=10$ symbols, and with switching with $T_s=2$ corresponding to the maximum switching rate. The results show that using an RIS with no switching garners the smallest KMR because it leads to better channel estimate due to the longer channel probing time, and still performs better than a system without an RIS because the added RIS channels improves the overall channel gains. On the other hand, an RIS with $T_{\rm s}=2$ shows the worst performance in terms of KMR because the number of channel probing time is just one symbol for each of Alice and Bob, leading to poor estimation accuracy and hence higher mismatch rate. However, with increasing $N$, the performances improves to rival the case with no RIS scenario due to the improved received signal corresponding to the increased number of reflectors.

In Fig. \ref{fig:qkmr}, we plot the KMR versus the signal-to-noise ratio (SNR) defined as $\text{SNR}=\frac{P}{\sigma^2}$, for different quantization levels $Q=2,4,8$. All considered simulation scenarios show a downwards trend as SNR increases which is expected since the channel estimation quality improves. Moreover, we can observe the effect of quantization on the KMR in the figure. The effect of $Q$ on the KMR can be seen implicitly in \eqref{matchprob} where the match probability $p$ decreases as $Q$ increases because the quantization resolution increases and it becomes more likely that quantized phases extracted at Alice and Bob do not match. This explains why the highest KMR in the figure is for a system with an RIS with $T_{\rm s}=2$ and $Q=8$.
{To summarize, KMR is affected primarily by $Q$ and $T_s$, increasing $T_s$ leads to lower KMR whilst increasing $Q$ leads to higher KMR.}
 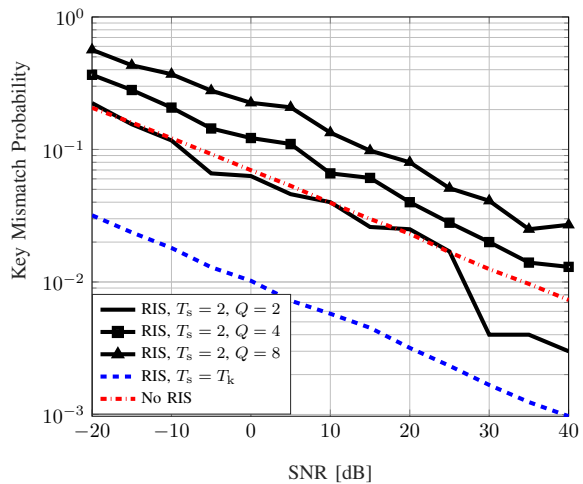
\begin{figure}[t]
     \label{fig:qkmr}
\tikzset{every picture/.style={scale=0.77}, every node/.style={scale=1}}
%
%
\begin{tikzpicture}

\begin{axis}[%
width=4.83in,
height=4.15in,
xmin=-20,
xmax=40,
xlabel style={at={(axis cs: 10,0.0008)}, font=\color{white!15!black}},
xlabel={SNR [dB]},
ymode=log,
ymin=0.000970000000000001,
ymax=1,
yminorticks=true,
ylabel style={font=\color{white!15!black}},
ylabel={Key Mismatch Probability},
xmajorgrids,
ymajorgrids,
yminorgrids,
legend style={at={(axis cs: -20,0.001)}, anchor = south west, legend cell align=left, align=left, draw=white!15!black}
]
\addplot [color=black, line width=2.0pt]
  table[row sep=crcr]{%
-20	0.224\\
-15	0.155\\
-10	0.117\\
-5	0.066\\
0	0.063\\
5	0.046\\
10	0.04\\
15	0.026\\
20	0.025\\
25	0.017\\
30	0.004\\
35	0.004\\
40	0.003\\
};
\addlegendentry{\footnotesize RIS, $T_{\rm s}=2$, $Q=2$}

\addplot [color=black, line width=2.0pt, mark=square, mark options={solid, black}]
  table[row sep=crcr]{%
-20	0.366\\
-15	0.281\\
-10	0.207\\
-5	0.144\\
0	0.122\\
5	0.11\\
10	0.066\\
15	0.061\\
20	0.04\\
25	0.028\\
30	0.02\\
35	0.014\\
40	0.013\\
};
\addlegendentry{\footnotesize RIS, $T_{\rm s}=2$, $Q=4$}

\addplot [color=black, line width=2.0pt, mark=triangle, mark options={solid, black}]
  table[row sep=crcr]{%
-20	0.565\\
-15	0.432\\
-10	0.37\\
-5	0.278\\
0	0.225\\
5	0.208\\
10	0.134\\
15	0.098\\
20	0.08\\
25	0.051\\
30	0.041\\
35	0.025\\
40	0.027\\
};
\addlegendentry{\footnotesize RIS, $T_{\rm s}=2$, $Q=8$}

\addplot [color=blue, dashed, line width=2.0pt]
  table[row sep=crcr]{%
-20	0.03187\\
-15	0.02367\\
-10	0.0180699999999999\\
-5	0.0129199999999998\\
0	0.0102099999999999\\
5	0.00722999999999993\\
10	0.00575999999999995\\
15	0.00450999999999996\\
20	0.00317999999999998\\
25	0.00233\\
30	0.00167\\
35	0.00124\\
40	0.000970000000000001\\
};
\addlegendentry{\footnotesize RIS, $T_{\rm s}=T_{\rm k}$}

\addplot [color=red, dashdotted, line width=2.0pt]
  table[row sep=crcr]{%
-20	0.20635\\
-15	0.15965\\
-10	0.12161\\
-5	0.0923799999999999\\
0	0.06975\\
5	0.0532000000000001\\
10	0.03963\\
15	0.02991\\
20	0.0231199999999999\\
25	0.0167799999999998\\
30	0.0124999999999999\\
35	0.0096699999999999\\
40	0.00731999999999992\\
};
\addlegendentry{\footnotesize No RIS}

\end{axis}

\end{tikzpicture}%
    \caption{Key mismatch rate versus SNR for different quantization levels. }
    \label{fig:qkmr}
\end{figure}

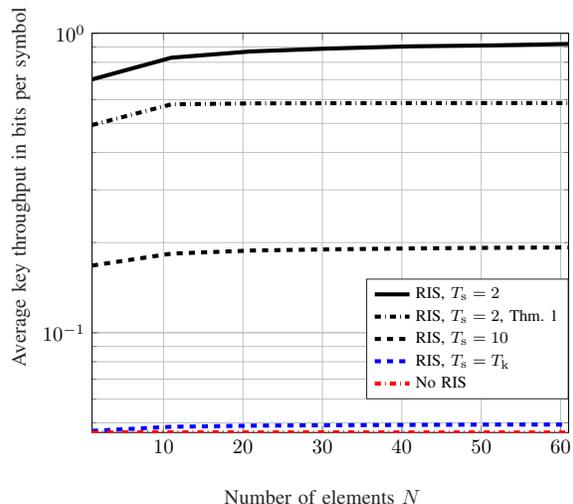
\begin{figure}[t]
\tikzset{every picture/.style={scale=0.77}, every node/.style={scale=1}}
%
%
\begin{tikzpicture}

\begin{axis}[%
width=4.83in,
height=4.15in,
xmin=1,
xmax=61,
xlabel style={at = {(axis cs: 30,0.04)}, font=\color{white!15!black}},
xlabel={Number of elements $N$},
ymode=log,
ymin=0.046424,
ymax=1,
yminorticks=true,
ylabel style={font=\color{white!15!black}},
ylabel={Average key throughput in bits per symbol},
xmajorgrids,
ymajorgrids,
yminorgrids,
legend style={at = {(axis cs: 61,0.06)}, anchor = south east,  legend cell align=left, align=left, draw=white!15!black}
]
\addplot [color=black, line width=2.0pt]
  table[row sep=crcr]{%
1	0.701299999999999\\
11	0.82873\\
21	0.86962\\
31	0.88934\\
41	0.90421\\
51	0.91056\\
61	0.92123\\
};
\addlegendentry{\footnotesize RIS, $T_{\rm s}=2$}

\addplot [color=black, dashdotted, line width=2.0pt]
  table[row sep=crcr]{%
1	0.493741202701072\\
11	0.579133247629705\\
21	0.583082239804802\\
31	0.584046618164438\\
41	0.584422414014336\\
51	0.584606752348685\\
61	0.584710618292995\\
};
\addlegendentry{\footnotesize RIS, $T_{\rm s}=2$, Thm. 1}

\addplot [color=black, dashed, line width=2.0pt]
  table[row sep=crcr]{%
1	0.167592\\
11	0.183718\\
21	0.188126\\
31	0.189888\\
41	0.191198\\
51	0.192114\\
61	0.192648\\
};
\addlegendentry{\footnotesize RIS, $T_{\rm s}=10$}

\addplot [color=blue, dashed, line width=2.0pt]
  table[row sep=crcr]{%
1	0.0470235\\
11	0.0485225\\
21	0.048922\\
31	0.0490915\\
41	0.0492195\\
51	0.049331\\
61	0.049351\\
};
\addlegendentry{\footnotesize RIS, $T_{\rm s}=T_{\rm k}$}

\addplot [color=red, dashdotted, line width=2.0pt]
  table[row sep=crcr]{%
1	0.0464425\\
11	0.0465065\\
21	0.04645\\
31	0.046492\\
41	0.04645\\
51	0.046455\\
61	0.046424\\
};
\addlegendentry{\footnotesize No RIS}

\end{axis}
\end{tikzpicture}%
    \caption{Average key throughput versus $N$. }
    \label{fig:avth}
\end{figure}

The average key throughput $\bar{R}_{\rm k}$ defined in \eqref{avgthrough} is depicted in Fig. \ref{fig:avth} versus the number of RIS elements $N$. We see that a system with an RIS with $T_{\rm s}=T_{\rm k}$ has the lowest throughput and is comparable with the case of no RIS in the system. However, despite the higher KMR for the cases with $T_{\rm s}=2$ and $10$, their average key throughput is higher. In the same figure, we plot the theoretical secret key rate lower bound $R_{\rm k}^{\rm lb}$ given in \eqref{rkmlb} for the case with RIS $T_{\rm s}=2$ in order to compare it against the experimental throughput. In all cases, the average key throughput increases initially with increasing $N$ due to the dependence of channel gains on $N$, and then stagnates after a certain $N$ value. { This is because the average key throughput depends on $N$ through the match probability $p$. After a certain $N$ value, $p$ equals one and there is no gain in increasing $N$ further. }

 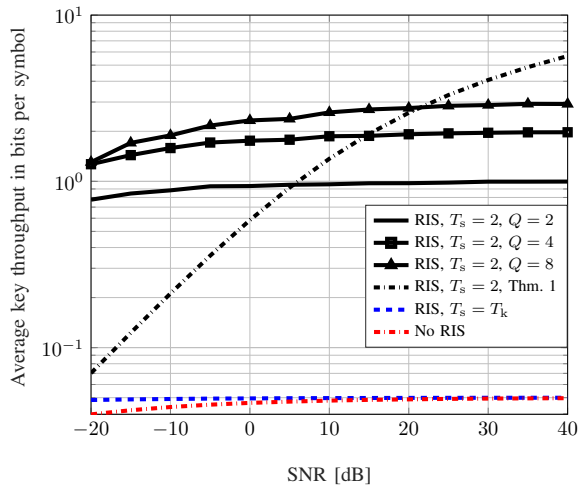
\begin{figure}[t]
\tikzset{every picture/.style={scale=0.77}, every node/.style={scale=1}}
%
%
\begin{tikzpicture}

\begin{axis}[%
width=4.83in,
height=4.15in,
xmin=-20,
xmax=40,
xlabel style={at={(axis cs: 10,0.032)}, font=\color{white!15!black}},
xlabel={SNR [dB]},
ymode=log,
ymin=0.0396825,
ymax=10,
yminorticks=true,
ylabel style={font=\color{white!15!black}},
ylabel={Average key throughput in bits per symbol},
xmajorgrids,
ymajorgrids,
yminorgrids,
legend style={at={(axis cs: 40,0.1)}, anchor = south east,  legend cell align=left, align=left, draw=white!15!black}
]
\addplot [color=black, line width=2.0pt]
  table[row sep=crcr]{%
-20	0.776\\
-15	0.845\\
-10	0.883\\
-5	0.934\\
0	0.937\\
5	0.954\\
10	0.96\\
15	0.974\\
20	0.975\\
25	0.983\\
30	0.996\\
35	0.996\\
40	0.997\\
};
\addlegendentry{\footnotesize RIS, $T_{\rm s}=2$, $Q=2$}

\addplot [color=black, line width=2.0pt, mark=square, mark options={solid, black}]
  table[row sep=crcr]{%
-20	1.268\\
-15	1.438\\
-10	1.586\\
-5	1.712\\
0	1.756\\
5	1.78\\
10	1.868\\
15	1.878\\
20	1.92\\
25	1.944\\
30	1.96\\
35	1.972\\
40	1.974\\
};
\addlegendentry{\footnotesize RIS, $T_{\rm s}=2$, $Q=4$}

\addplot [color=black, line width=2.0pt, mark=triangle, mark options={solid, black}]
  table[row sep=crcr]{%
-20	1.305\\
-15	1.704\\
-10	1.89\\
-5	2.166\\
0	2.325\\
5	2.376\\
10	2.598\\
15	2.706\\
20	2.76\\
25	2.847\\
30	2.877\\
35	2.925\\
40	2.919\\
};
\addlegendentry{\footnotesize RIS, $T_{\rm s}=2$, $Q=8$}

\addplot [color=black, dashdotted, line width=2.0pt]
  table[row sep=crcr]{%
-20	0.0702521960970939\\
-15	0.122757632055214\\
-10	0.211652628122156\\
-5	0.35735009554937\\
0	0.584866337079002\\
5	0.917653034101216\\
10	1.36795184955749\\
15	1.93039923599984\\
20	2.58493845608433\\
25	3.30615971479751\\
30	4.07135838192011\\
35	4.86378814170697\\
40	5.67242251316057\\
};
\addlegendentry{\footnotesize RIS, $T_{\rm s}=2$, Thm. 1}

\addplot [color=blue, dashed, line width=2.0pt]
  table[row sep=crcr]{%
-20	0.0484065\\
-15	0.0488165\\
-10	0.0490965\\
-5	0.049354\\
0	0.0494895\\
5	0.0496385\\
10	0.049712\\
15	0.0497745\\
20	0.049841\\
25	0.0498835\\
30	0.0499165\\
35	0.049938\\
40	0.0499515\\
};
\addlegendentry{\footnotesize RIS, $T_{\rm s}=T_{\rm k}$}

\addplot [color=red, dashdotted, line width=2.0pt]
  table[row sep=crcr]{%
-20	0.0396825\\
-15	0.0420175\\
-10	0.0439195\\
-5	0.045381\\
0	0.0465125\\
5	0.04734\\
10	0.0480185\\
15	0.0485045\\
20	0.048844\\
25	0.049161\\
30	0.049375\\
35	0.0495165\\
40	0.049634\\
};
\addlegendentry{\footnotesize No RIS}

\end{axis}

\end{tikzpicture}%
    \caption{Average key throughput versus SNR.}
    \label{fig:avgthrouq}
\end{figure}

Finally, Fig. \ref{fig:avgthrouq} shows the average throughput versus SNR, for different quantization levels. Note that $Q$ has an implicit adverse effect on $ \bar{R}_{\rm k}$ through $p$ as well as an explicit positive effect manifested by the increase of $\log_2(Q)$ with $Q$. We see that the $\log_2(Q)$ term dominates in the expression as evident by the RIS with  $T_{\rm s}=2,\ Q=8$ having the best average key throughput. Note that at certain high SNR, the theoretical secret key rate lower bound  $R_{\rm k}^{\rm lb}$ exceeds the experimental throughputs. This is because the simulated protocol only uses the channel phases and neglects channel amplitudes, whereas both are used in the derivation of the theoretical lower bound on $R_{\rm k}^{\rm lb}$.

\section{Conclusion}
 In this paper, we study the RIS effect in enhancing key generation, where the RIS provides a two-fold enhancement by adding additional channels and by perturbing the static channel in order to obtain a higher key rate. We formulate an expression for the theoretical achievable SKR lower bound using our proposed system model under block fading channels. Moreover, we derive the average key throughput for a proposed protocol and further study the effect of changing RIS parameters such as the number of elements $N$ and the switching rate of the RIS $T_s$, as well as system parameters such as the quantization $Q$ on the key throughput. {Using theoretical findings and simulations, we discover that increasing $N$ and $Q$ while decreasing $T_s$ yields the highest average key throughput.}  Future directions and extensions include finding the optimal parameters for the key throughput  {and investigating other channel models such as Ricean fading.}

\bibliographystyle{IEEEtran}
\bibliography{bib.bib}

\end{document}